\magnification= 1095
\headline={\ifnum\pageno>1 \hfil -- \folio\ -- \hfil \else\hfil\fi}
\footline={}

\raggedbottom
\overfullrule=0pt
\settabs 8 \columns\
\font\apj=cmcsc10
\font\smc cmcsc10 at 12 truept
\font\lge cmr12 at 14 truept
 at 14 truept
 at 10 truept

\def\ii{{\smc~ii}}

\def\posxr1{(J2000) $3^{\rm h}38^{\rm m}12.\!^{\rm s}29,
-24^{\circ}43^{\prime}48.\!^{\prime\prime}9$}

\def\posxr2{(J2000) $3^{\rm h}38^{\rm m}12.\!^{\rm s}66,
-24^{\circ}43^{\prime}47.\!^{\prime\prime}7$}

\def\posbl{(J2000) $3^{\rm h}38^{\rm m}12.\!^{\rm s}50,
-24^{\circ}43^{\prime}50.\!^{\prime\prime}3$}

\def\posgal{(J2000) $3^{\rm h}38^{\rm m}12.\!^{\rm s}39,
-24^{\circ}45^{\prime}07.\!^{\prime\prime}8$}

\def\posnvss{(J2000) $3^{\rm h}38^{\rm m}12.\!^{\rm s}93,
-24^{\circ}43^{\prime}51.\!^{\prime\prime}9$}

\def\obj{E~0336--248}
\def\ein{{\it Einstein}}
\def\ro{{\it ROSAT\/}}

\def\ref{\par\hangindent 25 pt\noindent}
\def\gsim{\lower 2pt \hbox{$\, \buildrel {\scriptstyle >}\over {\scriptstyle
\sim}\,$}}
\def\lsim{\lower 2pt \hbox{$\, \buildrel {\scriptstyle <}\over {\scriptstyle
\sim}\,$}}

\input psfig
\font\bigsf=cmssbx10 scaled 1400
\hbox{
\psfig{figure=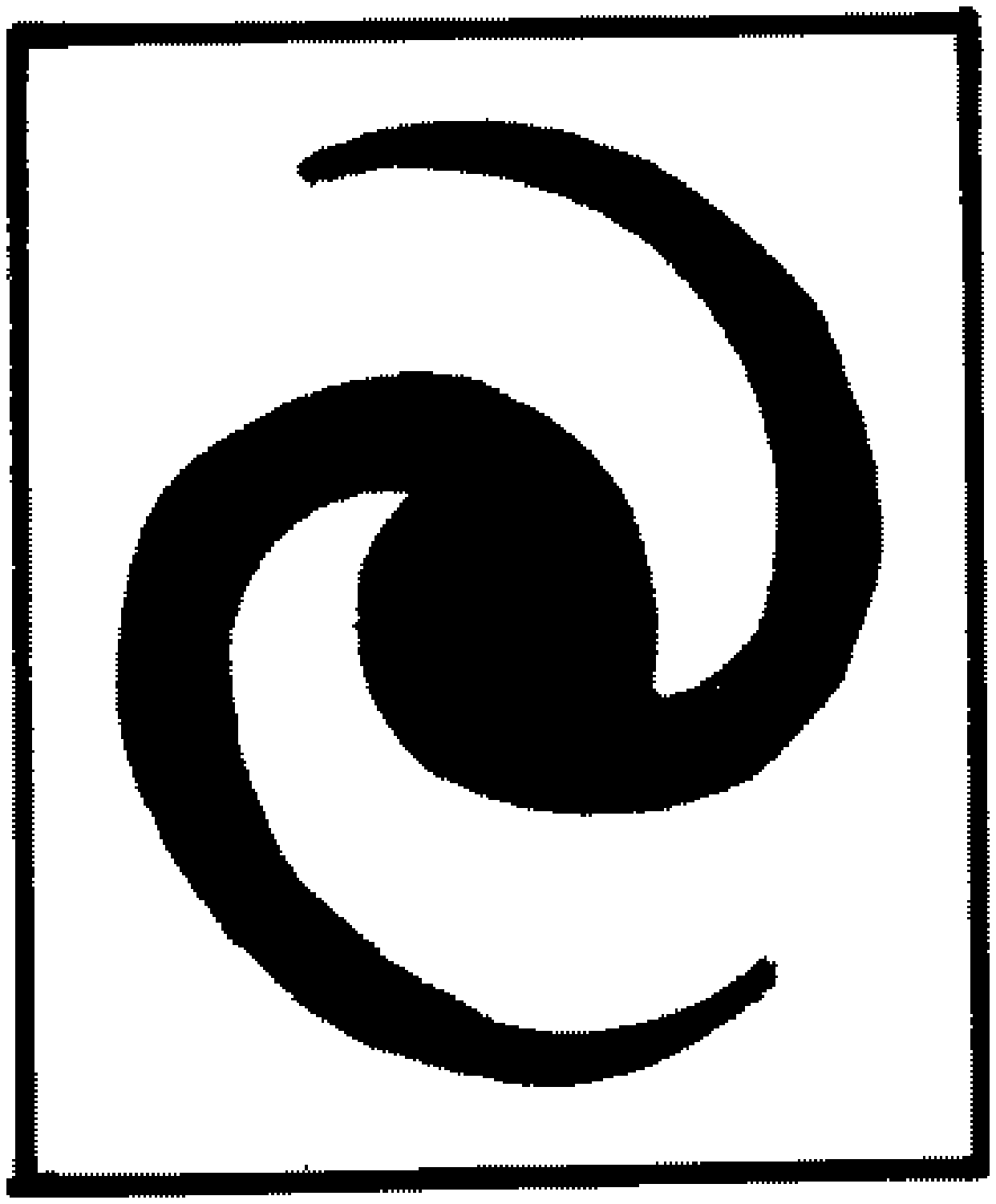,height=1in,width=1in}
\hskip 0.5truein 
\vbox to 1. truein {\bigsf \vfill
\noindent UNIVERSITY OF CALIFORNIA AT BERKELEY 
\medskip \noindent ASTRONOMY DEPARTMENT
\vfill}
}
\vfill

\font\lge cmr12 at 16 truept
\font\lit cmmi12 at 16 truept

\centerline{\lge E 0336--248 : A NEW BL LAC OBJECT}
\medskip
\centerline{\lge FOUND BY AN OLD {\lit EINSTEIN}}
\vfill

\centerline {\apj Jules P. Halpern$^{1,2}$ and Michael Eracleous$^{2,3}$}
\medskip
\centerline {Department of Astronomy, University of California, Berkeley, 
CA 94720}
\bigskip
\centerline {\apj Karl Forster}
\medskip
\centerline {Department of Astronomy, Columbia University, 
550 West 120th Street, New York, NY 10027}

\footnote{}{
\item{$^1$}
Permanent address: Department of Astronomy, Columbia University,
550 West 120th Street, New York, NY 10027.
\smallskip
\item{$^2$}
Visiting Astronomer, Kitt Peak National Observatory, National
Optical Astronomy Observatories, which is operated by AURA, Inc.,
under a cooperative agreement with the National Science Foundation.
\smallskip
\item{$^3$} Hubble Fellow. \smallskip
}

\vfill

\bigskip\bigskip
\centerline {To appear in {\it The Astronomical Journal}}

\vfill

\centerline {ABSTRACT}
\bigskip
\noindent
We obtained new \ro\ HRI and optical observations in the field of the
\ein\ X-ray source \obj , which we use to identify it as a 19th
magnitude BL~Lacertae object at $z = 0.251$ with $L_{\rm X} = 1 \times
10^{45}$ erg~s$^{-1}$.  It is also a 14~mJy radio source at 20~cm.  An
emission-line galaxy at $z = 0.043$ that was previously considered a
Seyfert identification for \obj\ is shown instead to be an unrelated,
non-active H~II region galaxy that lies $78^{\prime\prime}$ from the
X-ray source.  The resolution of this historical case of mistaken
identity illustrates that discoveries of non-AGN emission-line
galaxies with high X-ray luminosity should be tested carefully.  The
properties of \obj\ are similar to those of other X-ray selected
BL~Lacs, including its location in an apparent group or cluster of
galaxies.  Somewhat unusual is the weak contribution of nonstellar
optical light relative to the starlight in the spectrum of its host
galaxy, which raises once again the possibility that even
high-luminosity BL~Lac objects may be difficult to identify in X-ray
selected samples.  We discuss a possible manifestation of this problem
that appeared in the recent literature.

\vfill

\eject
\centerline {\apj 1. introduction}
\bigskip

Pravdo \& Marshall (1984) classified the serendipitous
\ein\ IPC X-ray source \obj\ as a narrow-line AGN at $z=0.043$
on the basis of a low-resolution optical spectrum that they obtained
of an anonymous galaxy near the X-ray position.  Consequently, this
object appears in standard AGN catalogs (Veron-Cetty \& Veron 1996,
Hewitt \& Burbidge 1983), as well as in a list of bright quasars to be
used in absorption-line studies with the {\it HST} (Bowen et
al. 1994).  Although bright enough, \obj\ was not included in the
\ein\ Medium Sensitivity Survey (Stocke et al. 1991) because it fell
close to a window support rib of the IPC detector.  We became
suspicious of the optical identification of \obj\ when our own higher
resolution optical spectra, taken for the purpose of studying the
nature of narrow-line X-ray galaxies, revealed only H~II region
emission lines, and no AGN features.  The implied X-ray luminosity of
$\approx 3 \times 10^{43}$~erg~s$^{-1}$ would be far in excess of that
observed from any non-AGN starforming galaxy.  We concluded that
either \obj\ is a highly unusual galaxy, or more likely, the
identification of it with the X-ray source is incorrect.  Meanwhile,
another serendipitous observation of \obj\ had been made, this time
with the \ro\ PSPC, but since both the IPC and the PSPC X-ray
positions were obtained far off-axis in their respective detectors,
neither was reliable enough to establish the identification
unambiguously.  Therefore, we obtained an additional X-ray observation
with the \ro\ HRI, enabling us to make a firm identification with a
much fainter optical object that turned out to be a BL~Lac object at
$z=0.251$ as measured from stellar absorption features in its optical
spectrum.  In the remainder of this paper, we describe the optical and
X-ray data collected in the course of this investigation, and discuss
some of the implications of the outcome.

\bigskip
\centerline{\apj 2. the wrong galaxy}
\bigskip

We obtained our first optical spectrum of the galaxy near
\obj\ from Bob Becker, who kindly observed it for us with the lens-grism
spectrograph at Lick Observatory on 1988 September 13.  As shown in
Figure~1a, this spectrum reveals no AGN features.  The absence of any
[O~III]$\lambda$5007 emission, despite the presence of weak H$\beta$,
raised serious doubts about its classification as an AGN.  (The slight
broadening of H$\beta$ is due to poor focus in the spectrograph, and
is therefore not indicative of AGN activity.)  Since the spectrum
obtained by Pravdo \& Marshall (1984) was of even lower resolution
than this one, we considered it likely that their Seyfert
classification was mistaken. Nevertheless, we decided to obtain a
higher resolution spectrum to look for weak, broad H$\alpha$ emission
that is sometimes the only evidence of an active nucleus in X-ray
selected AGNs, before declaring there to be a problem with the
identification.

Accordingly, we observed the galaxy with the Goldcam CCD spectrometer
on the KPNO 2.1m telescope on 1995 January 23.  A resolution of
4.2~\AA\ was achieved using a 600 lines mm$^{-1}$ grating blazed at
6750~\AA\ and a slit width of $1.\!^{\prime\prime}8$.  The wavelength
range covered was 5500--8500~\AA .  The heliocentric redshift that we
measure from this spectrum is $0.04255 \pm 0.00005$.  The portion of
the spectrum containing the H$\alpha$ line is shown in Figure~1b.
There are decidedly no AGN features in this spectrum.  The velocity
widths of H$\alpha$ and [N~II] are consistent with the instrumental
resolution, yielding a conservative upper limit on their FWHM of
80~km~s$^{-1}$.  The [N~II]$\lambda$6583/H$\alpha$ ratio is 0.37,
typical of H~II region galaxies.  There is no additional broad
component of H$\alpha$ present.  Despite the glaring absence of AGN
features in this particular spectrum, the possibility that it might
contain a highly variable AGN in a quiescent phase could not be
rejected out of hand.  Nor could we evaluate the possibility that the
galaxy makes a partial contribution to a confused or extended X-ray
source.  Therefore, the definitive test would have to come from a high
resolution X-ray observation with the \ro\ HRI.

\topinsert
\centerline{\hbox to 6.5truein{
\hskip -0.2 truein
\psfig{figure=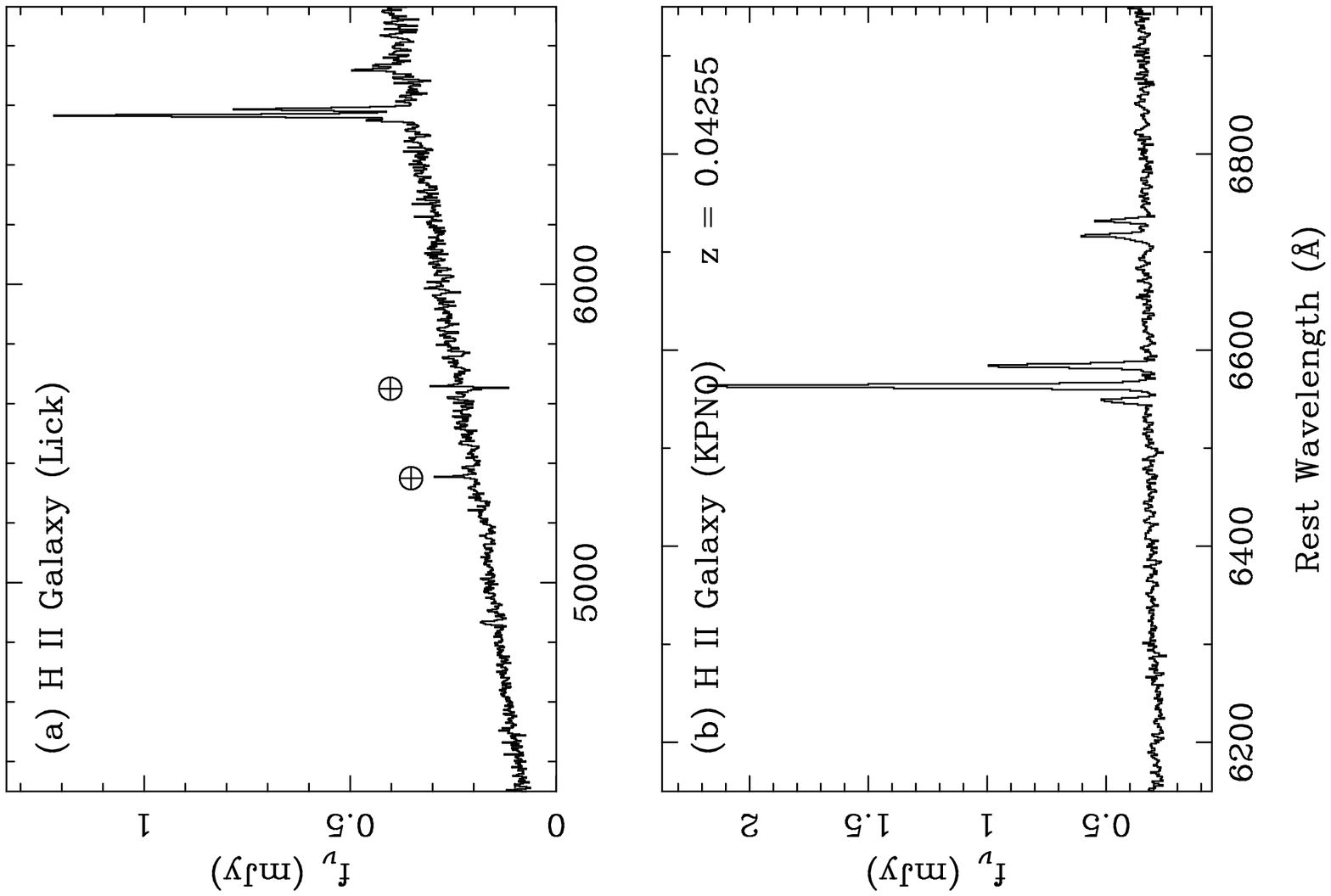,height=5in,rheight=4.2in,angle=-90}
\hskip -0.7 truein
\psfig{figure=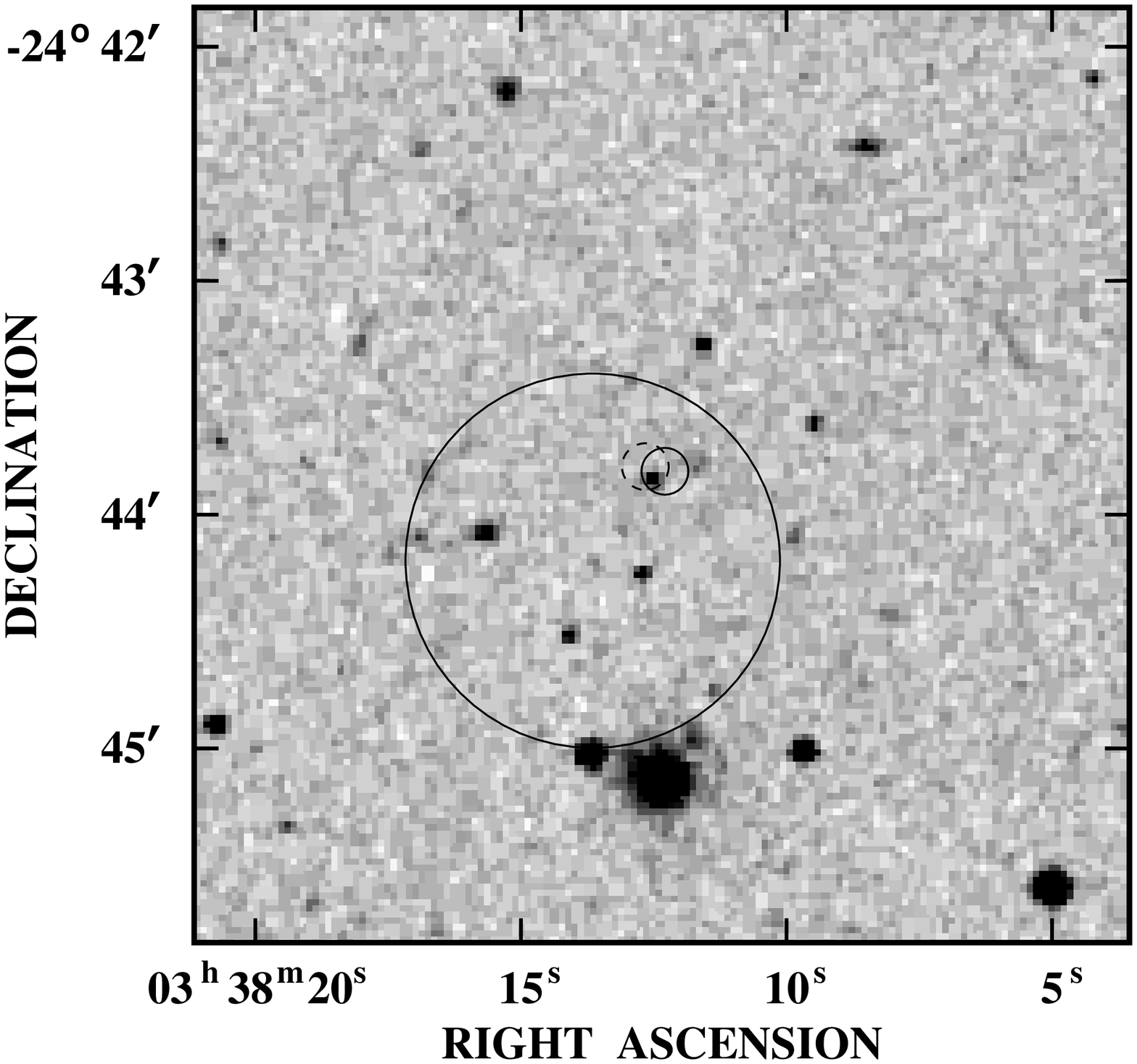,height=3in,rheight=3.5in,rwidth=3.5in}
}}
\bigskip
\centerline{\hbox to 6.5truein{
\vbox to 2 truein {\hsize=2.7 truein
\noindent {\apj Figure 1. --} 
Two spectra of the previously believed counterpart of \obj , shifted
to its rest frame.  Neither show any AGN features.  (a) Spectrum
obtained at Lick Observatory.  Note the absence of [O~III] $\lambda$5007 
or other Seyfert-like features.  (b) Higher resolution spectrum obtained 
on the KPNO 2.1m telescope.  The width of the emission lines is consistent 
with the instrumental resolution of 4.2~\AA . \vfill
}\hfill
\vbox to 2 truein {\hsize=3.3 truein
\noindent {\apj Figure 2. --} 
Finding chart for \obj\ from the digitized SERC Sky Survey IIIa-J
plate.  X-ray error circles from the \ein\ IPC and \ro\ HRI are shown;
the dashed circle is from the second HRI observation.  The optical
position of the BL Lac object within the HRI error circles is
(J2000) $3^{\rm h}38^{\rm m}12.\!^{\rm s}50$,
$-24^{\circ}43^{\prime}50.\!^{\prime\prime}3$. 
The unrelated H~II galaxy is the brightest object just south of the
IPC error circle, at (J2000) $3^{\rm h}38^{\rm m}12.\!^{\rm s}39$,
$-24^{\circ}45^{\prime}07.\!^{\prime\prime}8$. \vfill
}}}
\endinsert

\eject
\centerline {\apj 3. x-ray observations}
\bigskip
\centerline{\it 3.1 X-ray Position}
\medskip

We conducted a \ro\ HRI observation of \obj\ on 1996 Jan. 17 for
3616~s.  A single, pointlike X-ray source was detected at (J2000)
$3^{\rm h}38^{\rm m}12.\!^{\rm s}29,
-24^{\circ}43^{\prime}48.\!^{\prime\prime}9$ with a count rate of
$0.0497\pm 0.0038$~s$^{-1}$.  Figure~2 is a finding chart made from
the digitized SERC Southern Sky Survey IIIa-J plate, showing the \ro\
HRI error circle of radius $6^{\prime\prime}$, as well as the larger
\ein\ IPC error circle of radius $48^{\prime\prime}$.  The \ro\
position unambiguously determines the correct identification with a
$\approx 19$ mag object at \posbl .  In addition there is a 20~cm
radio source of flux density $14.0 \pm 1.1$~mJy at \posnvss , within
$6^{\prime\prime}$ of the optical object, in the NRAO VLA Sky Survey
(NVSS, Condon et al. 1997).  The positional difference is within the
expected error of that survey.

The bright galaxy previously thought to be the identification of the
X-ray source lies at \posgal , just south of the \ein\ IPC error
circle.  It is $78^{\prime\prime}$ from the actual X-ray source.
Since this galaxy is not detected in X-rays, we can place an upper
limit on its X-ray luminosity that is about a factor of 30 smaller
than previously thought, or less than $10^{42}$~erg~s$^{-1}$.
Virtually all H~II region galaxies fall below this limit (Moran,
Halpern, \& Helfand 1994, 1996; Halpern, Helfand, \& Moran 1995).  It
is also not detected in the NVSS.

Although the \ro\ HRI observation of \obj\ was entirely successful, an
error in the processing software mistakenly concluded otherwise, which
led to the automatic scheduling of an additional observation 7 months
later.  Despite our protestations that no more data were needed, a
second observation of 3262~s duration was carried out on 1996
Aug. 8--9.  The same source was detected at the overlapping position
\posxr2 , as shown by the dashed circle in Figure~2.  Thus, the
optical identification was confirmed, albeit with a reduced count rate
of of $0.0275 \pm 0.0030$~s$^{-1}$.  While it is comforting to know
that the X-ray source is variable in flux but not in position, it is
disconcerting that human intervention was not possible to prevent an
unintended \ro\ pointing from being made.

\medskip
\centerline{\it 3.2 X-ray Spectral Fitting}
\medskip

As mentioned above, \obj\ was also observed serendipitously
$36^{\prime}$ off-axis in a \ro\ PSPC observation of 49,500~s duration
on 1991 August 14-15.  In order to extract a reasonably reliable
spectrum, we used a circular extraction aperture of diameter
$12^{\prime}$, which is large enough to contain $\ge 95$\% of the
photons with energies below 2~keV for a point source at the given
off-axis distance (Hasinger et al. 1994). Background events were taken
from an aperture at the same off-axis distance and adjacent to the
target extraction aperture. This was done to avoid regions on the
detector that are shadowed by the PSPC window support structure.  The
light curve of the target and background were examined to exclude
periods of high background level, resulting in an accepted exposure
time of 39,048~s.

\topinsert
\centerline{\hbox to 6.5truein{
\vbox{\hsize= 3.1 truein
\psfig{figure=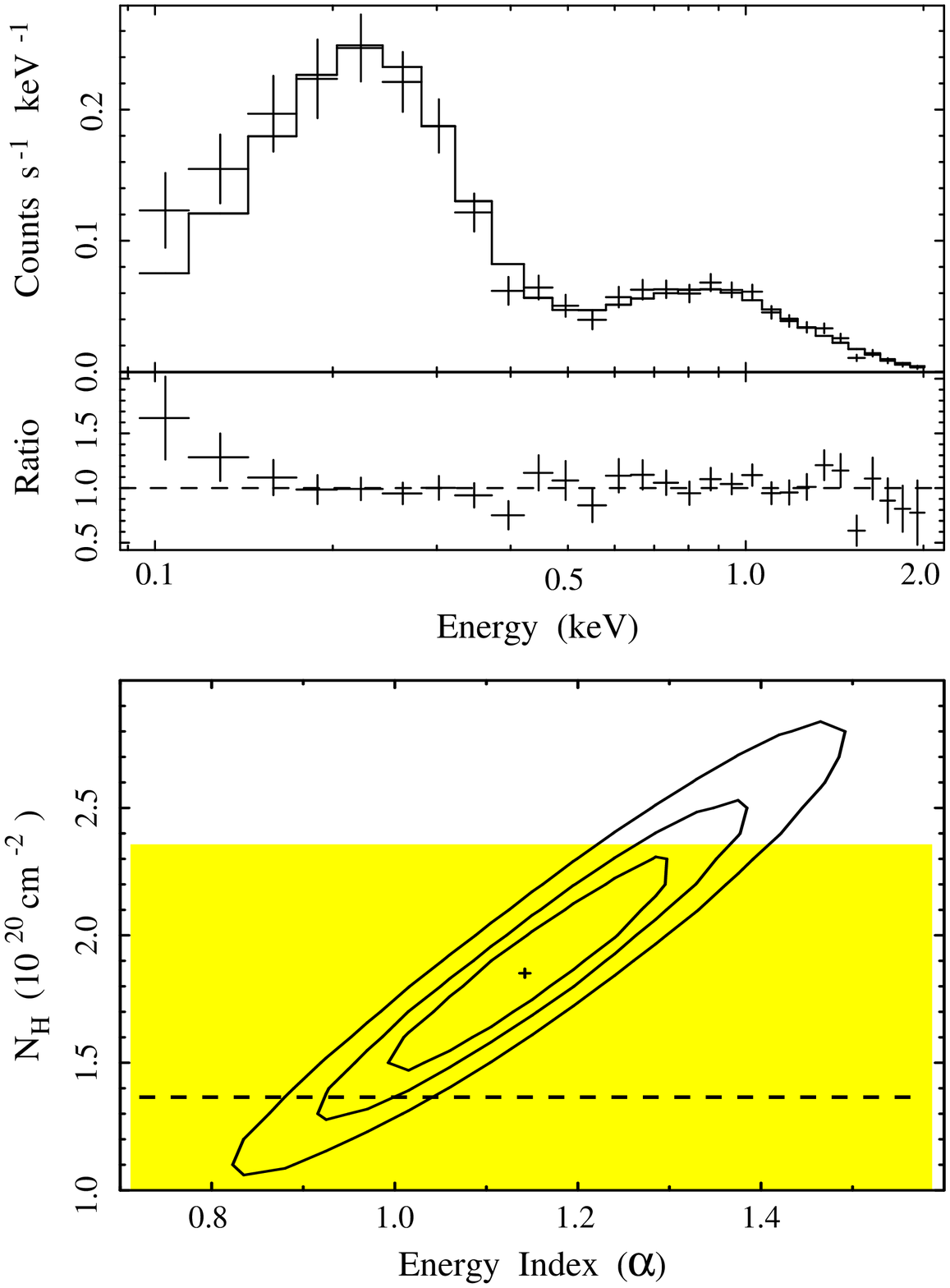,height=4.3in,rheight=5.in}
\medskip
\noindent {\apj Figure 3. --} 
The \ro\ PSPC spectrum of \obj\ showing the best fitting power-law
model and the data divided by the folded model.  Also shown are the
confidence contours of the power-law model fit.  These contours
represent the 68\%, 90\%, and 99\% confidence limits for two
interesting parameters.  The dashed line represents the Galactic
$N_{\rm H}$.
}\hfill 
\vbox {\hsize=2.9 truein \halign {
# \hfil & \hfil # \hfil  \cr
\multispan2 \hfil Table 1: The {\it ROSAT} PSPC \hfil \cr
\multispan2 \hfil Observation of E~0336--248 \hfil \cr
\noalign {\vskip 0.3cm \hrule \vskip 0.1cm \hrule \vskip 0.3cm}
\multispan2 \hfil Observational details \hfil \cr
\noalign {\vskip 0.3cm \hrule \vskip 0.3cm}
{$\alpha_{(2000)}$ \dotfill} & 
$03^{\rm h} \hskip 0.1cm 38^{\rm m} \hskip 0.1cm 12^{\rm s}.29$ \cr
{$\delta_{(2000)}$ \dotfill} & 
$-24^{\circ} \hskip 0.1cm 43^{\prime} \hskip 0.1cm 48^{\prime\prime}.9$ \cr
{$z$ \dotfill} & 0.2509 \cr 
{$N_{\rm H I}^{^{\rm Gal}}$ (cm$^{-2})\ ^{\rm a}$ \dotfill} & 
$(1.36\pm 1.0) \times 10^{20}$ \cr
{Date (UT) \dotfill} & 1991 Aug 14--15 \cr
{T$_{\rm obs}$ (s) \dotfill} & 39,048  \cr 
{Off-axis distance \dotfill} & 36$^{\prime}$ \cr
{Count rate $({\rm s}^{-1})\ ^{\rm b}$ \dotfill} & 0.112 $\pm$ 0.003 \cr
\noalign {\vskip 0.3cm \hrule \vskip 0.3cm}
\multispan2 \hfil Power-law model $^{\rm c}$ \hfil \cr
\noalign {\vskip 0.3cm \hrule \vskip 0.3cm}
{$\alpha$ \dotfill} & 1.14 $^{+0.26}_{-0.23}$ \cr
{$A$ ($\mu$Jy) $^{\rm d}$ \dotfill} & (0.449 $^{+0.045}_{-0.041}$) \cr
{$N_{\rm H}$ (cm$^{-2}$) \dotfill} & (1.85 $^{+0.69}_{-0.60}) 
\times 10^{20}$ \cr
{$\chi^{2}_{\nu}$ ($\nu$) \dotfill} & 1.111 (26) \cr
{$F_{\rm X}$ (erg cm$^{-2}$ s$^{-1})\ ^{\rm b}$ \dotfill} & 
(2.14 $^{+0.13}_{-0.11}) \times 10^{-12}$ \cr
{$L_{\rm X}$ (erg s$^{-1})\ ^{\rm e}$ \dotfill} & (1.14 $^{+0.44}_{-0.27})
\times 10^{45}$ \cr
\noalign {\vskip 0.3cm \hrule \vskip 0.1cm \hrule }
}
\medskip
\item{$^{\rm a}$}
Galactic neutral hydrogen column density from Stark et al. (1992).
\smallskip\item{$^{\rm b}$} 
0.1--2.0 keV observed frame, corrected for background and vignetting.
\smallskip\item{$^{\rm c}$} 
Errors are 90\% confidence for 2 interesting parameters.
\smallskip\item{$^{\rm d}$} 
Normalization at 1 keV observed frame.
\smallskip\item{$^{\rm e}$} 
0.1--2.0 keV rest frame ($H_0 = 50$ km s$^{-1}$ Mpc$^{-1},
q_0 = {1 \over 2} $), corrected for absorption.
}
}}
\bigskip
\endinsert

The PSPC response matrix is not reliable at energies above $\sim
2$~keV (Turner, Urry, \& Mushotzky 1993).  The following spectral
analysis is therefore limited to the range $0.1 < E < 2.0$ keV.  The
results of a simple power-law model fit to the observed spectrum are
given in Table~1 and shown in Figure~3 along with the $\chi^2$
confidence contours of the fit. The power-law model with energy index
$\alpha = 1.14 \pm 0.25$ is an adequate representation of the observed
spectrum, with $\chi^{2}_{\nu} = 1.111$ for 26 degrees of freedom, and
no significant improvement was achieved with other models.
The fitted column density, $N_{\rm H} = 1.85 \times10^{20}$~cm$^{-2}$,
is consistent with the Galactic value of $1.36 \times 10^{20}$~cm$^{-2}$ 
(Stark et al. 1992), considering that the uncertainty on the latter is
$\sim 1 \times 10^{20}$~cm$^{-2}$ (Elvis, Lockman, \& Fassnacht 1994).
Knowing that the correct redshift is 0.251 (see \S 4), the rest-frame
0.1--2.0~keV luminosity of \obj\ is $1.14 \times 10^{45}$ erg~s$^{-1}$
(H$_{0} = 50$ km s$^{-1}$ Mpc$^{-1}$, q$_{0} = {1 \over 2}$) after
correcting for absorption.  Among all four X-ray observations of
\obj , one by \ein\ and three by \ro , the observed 0.1--2.0~keV
flux ranges from $1.5 \times
10^{-12}$~erg~cm$^{-2}$~s$^{-1}$ to $3.3 \times
10^{-12}$~erg~cm$^{-2}$~s$^{-1}$.


\bigskip
\centerline{\apj 4. the correct optical identification}
\bigskip

Armed with the precise HRI position and the single coincident optical
object seen in Figure~3, we returned to Lick Observatory for
spectroscopic identification. The object was observed with the 3m
Shane reflector and Kast spectrograph (Miller \& Stone 1987) for
2400~s on 1996 October 11. The spectrum, which covers the range
4700--7470~\AA , was obtained through a $2^{\prime\prime}$ slit and
was reduced in a standard manner. The flux calibrated spectrum is
shown in Figure~4, after dereddening assuming a Seaton (1979) law and
color excess $E(B-V)=0.026$, and shifting to the rest frame of what is
obviously an old stellar population at $z = 0.2509 \pm 0.0005$.  The
resemblance of its spectrum to an elliptical galaxy, the absence of
any obvious emission lines, and the association with an x-ray source
of $L_{\rm X} \approx 1 \times 10^{45}$ erg~s$^{-1}$, all suggest that
\obj\ is a BL~Lac object.  The evidence for a nonstellar contribution
to the optical light (a hallmark of BL Lac objects) is definite but
subtle.  In comparison with normal elliptical galaxies such as
NGC~4339, the spectrum of \obj\ is bluer.  When the two spectra are
normalized at 6000~\AA\ as in Figure 4, the BL Lac spectrum is higher
by 40\% at 4000~\AA .  The Ca\ii\ H\&K line have larger equivalent
widths in the template galaxy, 23.5~\AA\ in NGC 4339 vs. 13.5~\AA\ in
\obj .

\topinsert
\centerline{\psfig{figure=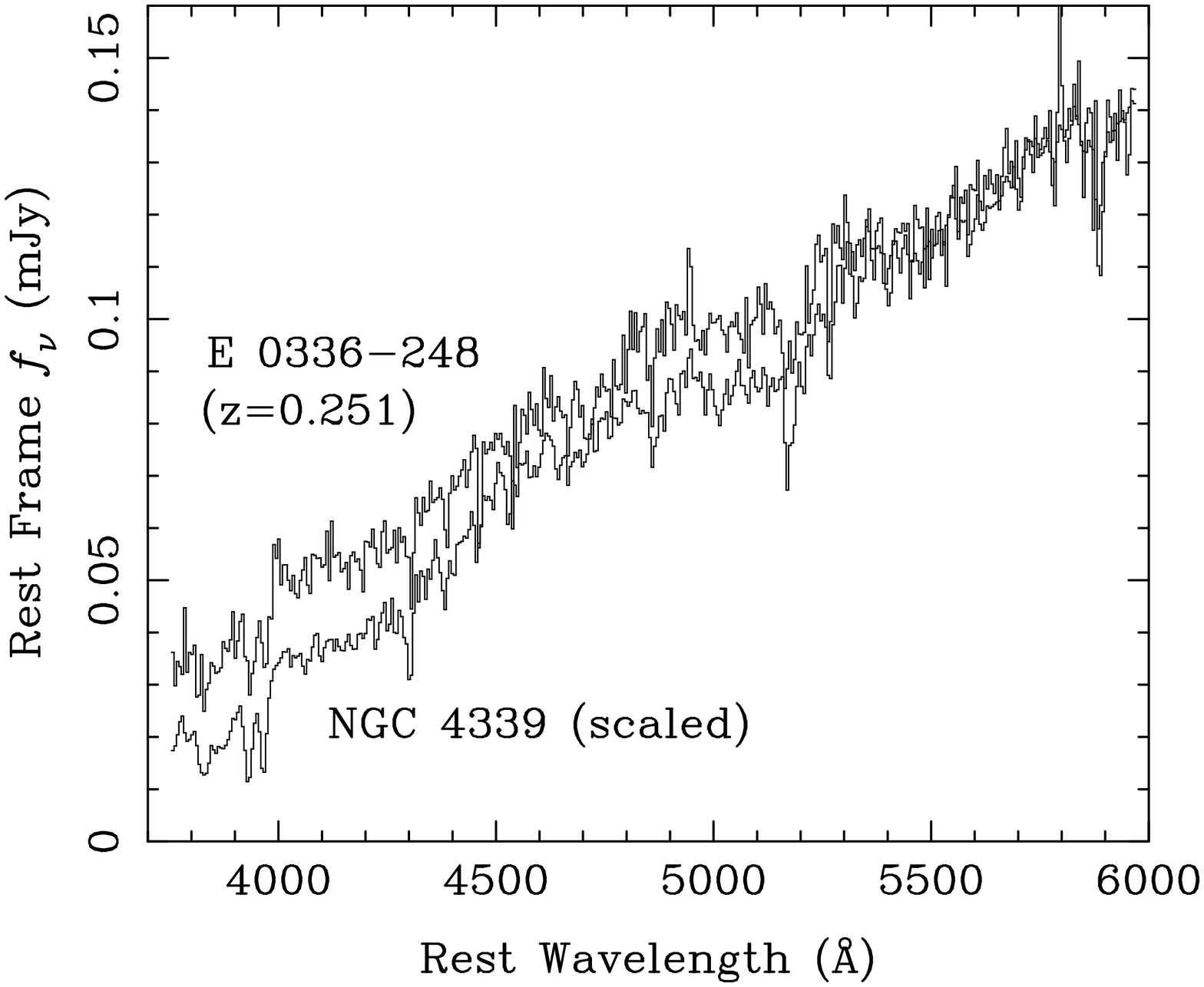,height=3in,rheight=3.5in,angle=-90}}
\centerline{\vbox{\hsize=5.5 truein \noindent {\apj Figure 4. --} 
Optical spectrum of the BL~Lac object identified with \obj , compared
with a spectrum of the elliptical galaxy NGC~4339.  The latter has
been normalized to \obj\ at 6000~\AA .
\medskip}}
\endinsert

In order to quantify the contribution of a featureless power law to
the optical spectrum, we fitted it with a model consisting of a linear
combination of starlight and a power law of the form $f_{\nu}\propto
\nu^{-\alpha}$.  The fit was carried out by exploring a range of
values of the power-law index in discrete steps, and determining the
linear combination coefficients at each step by minimizing the
r.m.s. deviation of the model from the data. To represent the
starlight we used the spectra of a number of elliptical galaxies as
templates. The galaxies used span a range of morphological types
(E0--E6) and include NGC 3379, NGC 4339, NGC 4365, NGC 5322, IC 4889,
and the composite giant elliptical galaxy of Yee \& Oke (1978). We
find that a reasonable fit can be achieved for all template galaxy
spectra and that a featureless continuum must be present because the
spectrum of \obj\ is flatter (bluer) than any of the template galaxy
spectra.  The decomposition into starlight and a featureless power-law
continuum yields a power-law index $\alpha=1.3\pm0.2$. The uncertainty
is systematic and is dominated by our lack of knowledge of the exact
stellar population of the host galaxy. The error bar quoted above
represents the dispersion in the power-law indices obtained using
different galaxy spectra as templates.  In the best-fitted
decomposition, the underlying galaxy contributes $(27\pm4)$\% of the
light just blueward of the Ca\ii\ H\&K break and $(45\pm4)$\% just
redward of it. In the $B$ and $V$ bands the galaxy contributes
$(55\pm3)$\% and $(63\pm3)$\% of the light, respectively.  Since the
galaxy contribution in the rest-frame $V$ band is $\approx 19.2$ mag,
its absolute $V$ magnitude is $-21.8$ at the luminosity distance of
1590~Mpc.  The galaxy is probably $\sim 0.3$ mag brighter than this,
or $\approx -22.1$, when account is taken of its light falling outside
the spectrograph slit.

\bigskip
\centerline {\apj 5. discussion and conclusions}
\bigskip

The observed properties of \obj\ are typical of those of other BL~Lac
objects discovered by \ein .  In particular, its X-ray spectral index
$\alpha$ is 1.1, and all but 1 of the 22 BL Lacs in the \ein\ Medium
Sensitivity Survey (EMSS) that were observed by \ro\ (Perlman et
al. 1996a) have $0.6 < \alpha < 1.9$.  Similarly, the broad-band
properties of \obj\ in radio, optical, and X-ray, place it in the
distinctive location populated by X-ray selected BL~Lacs in the
$(\alpha_{\rm ro},\alpha_{\rm ox})$ diagram (Stocke et al. 1990; Lamer
et al. 1996; Perlman et al. 1996b).  We estimate $\alpha_{\rm ro} =
0.45$ and $\alpha_{\rm ox} = 0.87$ using the total optical flux at
5500 \AA\ observed wavelength, the 6~cm radio flux extrapolated with a
flat spectrum to 20~cm, and the X-ray flux at 2~keV.

It is common for BL~Lac objects to be found in groups or clusters of
galaxies (Falomo 1996; Pesce, Falomo, \& Treves 1995). Around \obj ,
we can see faint images on the SERC and ESO Sky Survey plates which
may be an association of galaxies in which it resides.  One of these
objects, $25^{\prime\prime}$ south of the BL~Lac object in Figure 2,
is especially blue.  We obtained a spectrum of it which is of
insufficient quality to make a firm classification.  It has no strong
emission or absorption features.  In fact, we originally suspected it
as the X-ray source, but the HRI observations clearly show that this
is not the case.  Nevertheless, it would be interesting to find out
the nature of this neighboring object.

It is interesting to consider how the optical properties of the host
galaxy of \obj\ relate to the completeness of samples of X-ray
selected BL~Lac objects.  Because the fractional contribution of
nonstellar optical emission is relatively small in this object, its
spectrum differs from that of an ordinary elliptical galaxy only in
subtle ways.  In particular, the depth of the Ca\ii\ H\&K break, where
the flux drops by about 50\% in normal elliptical galaxies, is the
strongest indicator of dilution by nonstellar optical light.  In \obj
, the flux drops by 33\% at the break, which doesn't even meet
Stocke's nominal criterion of $< 25\%$ for classification as a BL~Lac
(Stocke et al. 1991; Morris et al. 1991).  This raises a concern about
the prospects of spectroscopic identification of BL~Lacs in more
luminous galaxies, because the absolute magnitude of the host galaxy
of \obj\ is only $M_V \approx -22.1$.  Host galaxies of BL~Lacs are
typically more luminous, having $\langle M_R \rangle$ in the range
$-23.1$ to $-23.5$ (Falomo 1996; Wurtz, Stocke, \& Yee 1996), and
total range $-21.7 < M_R < -24.4$.  Assuming $V-R=0.9$, this makes
\obj\ at $M_R \approx -23.0$ slightly less luminous than average.  If
its host galaxy were as little as 1 mag brighter, all other things
being equal, it would be virtually impossible to detect the dilution
by nonstellar light in ground-based spectroscopy.

Fortunately, radio emission supports the identification; there is
still no good evidence of the long-sought radio-quiet BL~Lacs.
However, there is seemingly justified recurrent speculation (e.g.,
Browne \& March\~a 1993) that a population of low-luminosity BL~Lacs
remains undiscovered in X-ray selected samples because of the
spectroscopic contrast problem described above.  Long ago, we
speculated that even highly luminous BL~Lacs might be misclassified as
either ``normal'' galaxies or cluster X-ray sources because their
optical nucleus can be camouflaged by a luminous host galaxy (Halpern
et al. 1986).  We revive that worry here, imagining \obj\ in a bigger
galaxy as a possible model.  In fact, the unusual galaxy J2310--43
recently described by Tananbaum et al. (1997) might be just such an
object.  It is less luminous than \obj\ by a factor $\approx 6-7$ in
X-rays and $\approx 2$ in radio, yet it is in a more luminous galaxy
of $M_R = -23.5$.  The principal obstacle to interpreting the origin
of its $> 10^{44}$ erg~s$^{-1}$ X-ray emission is the absence of
evidence of nuclear activity in the optical spectrum.  But if
J2310--43 hosts a BL Lac with the same nuclear spectral energy
distribution as \obj , then the dilution of its starlight spectrum by
its optical nonstellar continuum would be negligible, which seems to
be consistent with observations.  For this reason we regard the BL~Lac
hypothesis for J2310--43, considered by Tananbaum et al. (1997), as
quite plausible.

Finally, we repeat our plea for high-quality optical spectroscopy when
classifying narrow emission-line galaxies.  Unlike the H~II galaxy
misidentification reported here, two other ambiguous narrow-line
spectra from Pravdo \& Marshall (1984), E~0116+317 and E~1242+165,
were subsequently classified as bona-fide Seyfert galaxy X-ray sources
in the EMSS (Stocke et al. 1991).  The latter is by far the more
common outcome.  To paraphrase Halpern et al. (1995), there are still
no X-ray luminous starburst galaxies.

\medskip
We thank Bob Becker for helping to launch this study by obtaining a
spectrum for us at Lick Observatory.  J. P. H. and K. F. acknowledge
support from NASA ROSAT grant NAG~5-1935.  M. E. acknowledges support
from Hubble Fellowship grant HF-01608.01-94A, awarded by the Space
Telescope Science Institute, which is operated by AURA, Inc.  for NASA
under contract NAS 5-26555.  This paper is contribution 634 of the
Columbia Astrophysics Laboratory.


\bigskip
\centerline{REFERENCES}
\bigskip
\ref
Bowen, D.~V., Osmer, S.~J., Blades, C.~J., Tytler,~D., Cottrell,~L.,
Fan, X.-M., \& Lanzetta, K.~M. 1994, AJ, 107, 461
\ref
Browne, I., \& March\~a, M. 1993, MNRAS, 261, 795
\ref
Condon, J.~J., Cotton, W.~D., Greisen, E.~W., Yin, Q.~F,
Perley, R.~A., Taylor, G.~B. \&  Broderick, J.~J. 1997, preprint
\ref
Elvis, M., Lockman, F.J., \& Fassnacht, C. 1994, ApJS, 95, 413 
\ref
Falomo, R. 1996, MNRAS, 283, 241
\ref
Halpern, J.~P., Helfand, D.~J., \& Moran, E.~C. 1995, ApJ, 453, 611
\ref
Halpern, J.~P., Impey,~C., Bothun,~G., Tapia,~S., Skillman,~E.,
Wilson,~A., \& Meurs,~E. 1986, ApJ, 302, 711
\ref
Hasinger, G., Boese, G., Predehl, P., Turner, J.~T., Yusaf, R., 
George, I.~M., \& Rohrbach, G. 1994, MPE/OGIP Calibration memo CAL/ROS/93-015
\ref
Hewitt,~A., \& Burbidge,~G. 1993, ApJS, 87, 451
\ref
Lamer, G., Brunner, H., \& Staubert, R. 1996, A\&A, 311, 384
\ref
Miller, J.~S., \& Stone, R.~P.~S. 1987, Lick Obs. Tech. Rep., No. 48
\ref
Moran, E.~C., Halpern, J.~P., \& Helfand, D. J. 1994, ApJ, 433, L65
\ref
------------. 1996, ApJS, 106, 341
\ref
Morris, S.~L., Stocke, J.~T., Gioia, I.~M., Schild, R.~E., \&
Wolter, A. 1991, ApJ, 380, 49
\ref
Perlman, E.~S., Stocke, J.~T., Wang, Q.~D., \& Morris, S.~D.
1996a, ApJ, 456, 451
\ref
Perlman, E.~S., Stocke, J.~T., Schachter, J.~F., Elvis,~M., Ellingson,~E.,
Urry, C.~M., Potter,~M., \& Impey,, C.~D., 1996b, ApJS, 104, 251
\ref
Pesce, J.~E., Falomo, R., and Treves, A. 1995 AJ, 110, 1554
\ref
Pravdo, S.~H., \& Marshall, F.~E. 1984, ApJ, 281, 570
\ref
Seaton, M. J. 1979, MNRAS, 187, 73P
\ref
Stark, A.~A., Gammie, C.~F., Wilson, R.~W., Bally, J., Linke, R.~A., 
Heiles, C., \& Hurwitz, M. 1992, ApJS, 79, 77
\ref
Stocke, J.~T., Morris, S.~L., Gioia, I., Maccacaro, T., Schild, R.~E.,
\& Wolter, A. 1990, ApJ, 348, 141
\ref
Stocke, J.~T., Morris, S.~L., Gioia, I.~M., Maccacaro,~T.,
Schild,~R., Wolter,~A., Fleming, T.~A., \& Henry, J.~P. 1991, ApJS, 76, 813
\ref
Tananbaum, H., Tucker, W., Prestwich, A., \& Remillard, R. 1997, ApJ, 
476, 83
\ref
Turner, T.~J., Urry, C.~M., \& Mushotzky, R.~F. 1993, ApJ, 418, 653
\ref
Veron-Cetty, M.~P., \& Veron, P. 1996, ESO Special Report, 17, 1
\ref
Wurtz, R., Stocke, J.~T., \& Yee, H.~K.~C. 1996, ApJS, 103, 109
\ref
Yee, H.~K.~C., \& Oke, J.~B. 1978, ApJ, 226, 753
\vfill

\bye